\documentclass{LMCS}
\usepackage{amsmath}
\usepackage{amsfonts}
\usepackage{amssymb}
\usepackage{latexsym}
\usepackage{hyperref}
\usepackage[latin1]{inputenc}

\theoremstyle{plain}\newtheorem{fact}[thm]{Fact}

\newenvironment{proofofthm}[1]{\begin{trivlist} \item[\hspace{\labelsep}\em Proof of Theorem~\ref{#1}:]}{\qed\end{trivlist}}

\newcommand{\bfmid}{~~~\mathbf{|}~~~}
\renewcommand{\sl}{\em}
\newcommand{\Act}{A}
\newcommand{\Ev}{E}

\newcommand{\depth}{{\mathit{depth}}}

\newcommand{\goto}[1]{\stackrel{#1}{\longrightarrow}}   
\newcommand{\mv}[1]{\mathrel{\stackrel{#1}{\longrightarrow}}}

\newcommand{\nil}{\mathbf{0}}
\newcommand{\hmerge}{\mathrel{|^{\negmedspace\scriptstyle /}}}

\newcommand{\TwoRules}[4]{
\begin{displaymath}
\begin{array}{c}
#1 \\\hline 
#2
\end{array}
\qquad
\begin{array}{c}
#3 \\\hline 
#4
\end{array}
\end{displaymath}
}

\newcommand{\ThreeRules}[6]{
\begin{displaymath}
\begin{array}{c}
#1 \\\hline 
#2
\end{array}
\qquad
\begin{array}{c}
#3 \\\hline 
#4
\end{array}
\qquad
\begin{array}{c}
#5 \\\hline 
#6
\end{array}
\end{displaymath}
}

\newcommand{\bistext}{{\underline{\leftrightarrow}_{\it 2S}}}
\newcommand{\bis}{~{\underline{\leftrightarrow}_{\it 2S}}~}
\newcommand{\nbis}{~{\underline{\leftrightarrow}_{\it 2S}}\!\!\!\!\!\!\!\!\!\!/~~~~~}
\newcommand{\HCCS}{\text{CCS}_{\scriptstyle H}}

\def\doi{1 (1:3) 2005}
\lmcsheading%
{\doi}
{12}  
{}
{}
{Sep.~\phantom{0}1, 2004} 
{Mar.~\phantom{0}9, 2005} 
{}    

\begin{document}

\title[Split-2
   Bisimilarity has a Finite Axiomatization over CCS with Hennessy's
   Merge]{Split-2
   Bisimilarity has a Finite Axiomatization over CCS with Hennessy's
   Merge}

\author[L.~Aceto]{\href{http://www.cs.aau.dk/~luca/}{Luca Aceto\rsuper
a}}

\address{\href{http://www.brics.dk}{{\lsuper a}BRICS},
\href{http://www.cs.auc.dk/}{Department of Computer Science},
\href{http://www.auc.dk/}{Aalborg University}, 9220 Aalborg \O,
Denmark, and School of Computer Science,
\href{http://www.ru.is/}{Reykjav\'{\i}k University}, Iceland}

\email{luca@cs.aau.dk, luca@ru.is} 

\author[W.~Fokkink]{\href{http://www.cwi.nl/~wan}{Wan Fokkink\rsuper b}}     

\address{{\lsuper b}Vrije Universiteit Amsterdam,
  \href{http://www.cs.vu.nl/}{Department of Computer Science}, Section
  Theoretical Computer Science, De Boelelaan 1081a, 1081 HV Amsterdam,
  The Netherlands} 

\email{wanf@cs.vu.nl}  
 
\author[A.~Ingolfsdottir]{\href{http://www.cs.auc.dk/~annai/}{Anna
Ingolfsdottir\rsuper c}}     
\address{\href{http://www.brics.dk}{{\lsuper c}BRICS},
\href{http://www.cs.auc.dk/}{Department of Computer Science},
\href{http://www.auc.dk/}{Aalborg University}, 9220 Aalborg \O,
Denmark, and Department of Computer Science, University of Iceland,
Iceland}     
\email{annai@cs.aau.dk, annaing@hi.is}  

\author[B.~Luttik]{\href{http://www.win.tue.nl/~luttik/}{Bas Luttik\rsuper d}}     

\address{\href{http://www.win.tue.nl/}{{\lsuper d}Department of Mathematics and Computer Science}, 
  \href{http://www.tue.nl/}{Eindhoven Technical University}, 
  5600 MB Eindhoven,
  The Netherlands}     

\email{luttik@win.tue.nl}  


\begin{abstract}
  This note shows that split-2 bisimulation equivalence (also known as
  timed equivalence) affords a finite equational axiomatization over
  the process algebra obtained by adding an auxiliary operation
  proposed by Hennessy in 1981 to the recursion, relabelling and
  restriction free fragment of Milner's Calculus of Communicating
  Systems. Thus the addition of a single binary operation,
  viz.~Hennessy's merge, is sufficient for the finite equational
  axiomatization of parallel composition modulo this non-interleaving
  equivalence. This result is in sharp contrast to a theorem
  previously obtained by the same authors to the effect that the same
  language is not finitely based modulo bisimulation equivalence.
\end{abstract}

\maketitle

\section{Introduction}\label{Sect:introduction}

This note offers a contribution to the study of equational
characterizations of the parallel composition operation modulo
(variations on) the classic notion of bisimulation
equivalence~\cite{Mi89,Pa81}. In particular, we provide a finite
equational axiomatization of {\sl split-2 bisimulation
equi\-va\-lence}---a notion of bisimulation equivalence based on the
assumption that actions have observable beginnings and
endings~\cite{GV87,GorrieriLaneve1995,He88}---over the recursion,
relabelling and restriction free fragment of Milner's CCS~\cite{Mi89}
enriched with an auxiliary operator proposed by Hennessy in a 1981
preprint entitled {\sl ``On the relationship between time and
interleaving''} and its published version~\cite{He88}. To put this
contribution, and its significance, in its research context, we find
it appropriate to recall briefly some of the key results in the
history of the study of equational axiomatizations of parallel
composition in process algebra.

Research on equational axiomatizations of behavioural equivalences
over process algebras incorporating a notion of parallel composition
can be traced at least as far back as the seminal paper~\cite{HM85},
where Hennessy and Milner offered, amongst a wealth of other classic
results, a complete equational axiomatization of bisimulation
equivalence over the recursion free fragment of CCS. (See the
paper~\cite{Baeten2004} for a more detailed historical account
highlighting, e.g., Hans Beki\'{c}'s early contributions to this field
of research.) The axiomatization given by Hennessy and Milner in that
paper dealt with parallel composition using the so-called {\sl
expansion law}---an axiom schema with a countably infinite number of
instances that is essentially an equational formulation of the
Plotkin-style rules describing the operational semantics of parallel
composition. This raised the question of whether the parallel
composition operator could be axiomatized in bisimulation semantics by
means of a finite collection of equations.  This question was answered
positively by Bergstra and Klop, who gave in~\cite{BK84b} a finite
equational axiomatization of the merge operator in terms of the
auxiliary left merge and communication merge operators. Moller
clarified the key role played by the expansion law in the
axiomatization of parallel composition over CCS by showing
in~\cite{Mo89,Mo90,Mo90a} that strong bisimulation equivalence is {\sl
not} finitely based over CCS and PA without the left merge
operator. (The process algebra PA~\cite{BK84b} contains a parallel
composition operator based on pure interleaving without communication
and the left merge operator.)  Thus auxiliary operators like the ones
used by Bergstra and Klop are indeed necessary to obtain a finite
axiomatization of parallel composition.  Moreover, Moller proved
in~\cite{Mo89,Mo90} that his negative result holds true for each
``reasonable congruence'' that is included in standard bisimulation
equivalence. In particular, this theorem of Moller's applies to
split-2 bisimulation equivalence since that equivalence is
``reasonable'' in Moller's technical sense.

In his paper~\cite{He88}, Hennessy proposed an axiomatization of
observation congruence~\cite{HM85} (also known as rooted weak
bisimulation equivalence) and timed congruence (essentially rooted
weak split-2 bisimulation equivalence) over a CCS-like recursion,
relabelling and restriction free process language. Those
axiomatizations used an auxiliary operator, denoted $\hmerge$ by
Hennessy, that is essentially a combination of Bergstra and Klop's
left and communication merge operators.  Apart from having soundness
problems (see the reference~\cite{Ac94c} for a general discussion of
this problem, and corrected proofs of Hennessy's results), the
proposed axiomatization of observation congruence is {\sl infinite},
as it used a variant of the expansion theorem
from~\cite{HM85}. Confirming a conjecture by Bergstra and Klop
in~\cite[page~118]{BK84b}, and answering problem~8
in~\cite{Aceto2003a}, we showed in~\cite{AFIL2003} that the language
obtained by adding Hennessy's merge to CCS does {\sl not} afford a
finite equational axiomatization modulo bisimulation equivalence. This
is due to the fact that, in strong bisimulation semantics, no finite
collection of equations can express the interplay between interleaving
and communication that underlies the semantics of Hennessy's merge.
Technically, this is captured in our proof of the main result
in~\cite{AFIL2003} by showing that no finite collection of axioms that
are valid in bisimulation semantics can prove all of the equations in
the following family:
 \begin{eqnarray*}
  a\nil \hmerge \sum_{i=0}^{n} \bar{a} a^i & \approx & 
  a \sum_{i=0}^{n} \bar{a} a^i + \sum_{i=0}^{n} \tau a^i \quad (n\geq 0) \enspace .
  \end{eqnarray*}
  In split-2 semantics, however, these equations are not sound, since
  they express some form of interleaving. 
  Indeed, we prove that, in sharp contrast to the situation in
  standard bisimulation semantics, the language with Hennessy's merge
  {\sl can be finitely axiomatized} modulo split-2 bisimulation
  equivalence, and its use suffices to yield a finite axiomatization
  of the parallel composition operation. This shows that, in
  contrast to the results offered in~\cite{Mo89,Mo90}, ``reasonable
  congruences'' finer than standard bisimulation equivalence can be
  finitely axiomatized over CCS using Hennessy's merge as the single
  auxiliary operation---compare with the non-finite axiomatizability
  results for these congruences offered in~\cite{Mo89,Mo90}.

\medskip 

The paper is organized as follows. We begin by presenting
preliminaries on the language $\HCCS$---the extension of CCS with
Hennessy's merge operator---and split-2 bisimulation equivalence in
Sect.~\ref{Sect:HCCS}. We then offer a finite equational axiom system for
split-2 bisimulation equivalence over $\HCCS$, and prove that it is
sound and complete (Sect.~\ref{Sect:ax}). 

\medskip 


This is a companion paper to~\cite{AFIL2003}, where the interested
readers may find further motivation and more references to related
literature. However, we have striven to make it readable independently
of that paper. Some familiarity with~\cite{Ac94c,He88} and the basic
notions on process algebras and bisimulation equivalence will be
helpful, but is not necessary, in reading this study. The uninitiated
reader is referred to the textbooks~\cite{BW90,Mi89} for extensive
motivation and background on process algebras. Precise pointers to
material in~\cite{Ac94c,He88} will be given whenever necessary.
 
%
%

\section{The language $\HCCS$}\label{Sect:HCCS}

The language for processes we shall consider in this paper, henceforth
referred to as $\HCCS$, is obtained by adding Hennessy's merge
operator from~\cite{He88} to the recursion, restriction and
relabelling free subset of Milner's CCS~\cite{Mi89}. This language is
given by the following grammar:

\[
p ::= \nil \bfmid \mu p \bfmid p+p \bfmid p ~|~ p
\bfmid p \hmerge p \enspace ,
\]
where $\mu$ ranges over a set of {\sl actions} $\Act$. We assume that
$\Act$ has the form $\{\tau\}\cup \Lambda \cup \bar{\Lambda}$, where
$\Lambda$ is a given set of {\sl names}, $\bar{\Lambda}= \{ \bar{a}
\mid a\in \Lambda \}$ is the set of {\sl complement names}, and $\tau$ is a
distinguished action.  Following Milner~\cite{Mi89}, the action 
$\tau$ will result from the synchronized occurrence of the
complementary actions $a$ and $\bar{a}$. We let $a,b$ range over the
set of {\sl visible actions} $\Lambda \cup \bar{\Lambda}$.  As usual,
we postulate that $\bar{\bar{a}}=a$ for each name $a\in\Lambda$. We
shall use 
$p,q,r$ to range over process terms.  The
{\sl size} of a term is the number of operation symbols in it.
Following standard practice in the literature on CCS and related
languages, trailing {\bf 0}'s will often be omitted from terms.

The structural operational semantics for the language $\HCCS$ given by
Hennessy in Sect.~2.1 of~\cite{He88} is based upon the idea that visible
actions have a beginning and an ending. Moreover, for each visible
action $a$, these distinct events may be observed, and are denoted by
$S(a)$ and $F(a)$, respectively. We define
\[
\Ev = \Act \cup \{S(a),F(a) \mid a \in \Lambda \cup \bar{\Lambda} \} 
\enspace .
\]
In the terminology of~\cite{He88}, this is the set of {\sl events},
and we shall use $e$ to range over it. As usual, we write $ E^*$ for
the collection of finite sequences of events.

The operational semantics for the language $\HCCS$ is given in terms
of binary next-state relations $\mv{e}$, one for each event $e\in \Ev$.
As explained in~\cite{He88}, the relations $\mv{e}$ are defined over
the set of {\sl states} $S$, an extension of $\HCCS$ obtained by
adding new prefixing operations $a_{\scriptstyle S}$ ($a\in \Lambda \cup
\bar{\Lambda}$) to the signature for $\HCCS$.  More formally, the set
of states is given by the following grammar:
\[
s ::= p \bfmid a_{\scriptstyle S} p \bfmid s~|~ s
\enspace ,
\]
where $p$ ranges over $\HCCS$. Intuitively, a state of the form
$a_{\scriptstyle S} p$ is one in which the execution of action $a$ has
started, but has not terminated yet. We shall use $s,t$ to range over
the set of states $S$.
\begin{table}
\caption{SOS Rules for $S$ ($\mu\in
A$, $a\in \Lambda \cup
\bar{\Lambda}$ and $e\in\Ev$)}
\ThreeRules{}{ap \mv{S(a)} a_{\scriptstyle S} p}
           {}{a_{\scriptstyle S} p \mv{F(a)}p}
           {}{\mu p \mv{\mu} p}
\TwoRules{p\mv{e} s}{p+q\mv{e} s}
           {q\mv{e} s}{p+q\mv{e} s}

\ThreeRules{s\mv{e} s'}{s~|~t\mv{e} s'~|~t}
           {t\mv{e} t'}{s ~|~t\mv{e} s~|~t'}
{s\mv{a} s',~t\mv{\bar{a}} t'}{s ~|~t\mv{\tau} s'~|~t'}

\TwoRules{p\mv{e} s}{p\hmerge q \mv{e} s~|~q}
{p\mv{a} p',~q\mv{\bar{a}} q'}{p \hmerge q\mv{\tau} p'~|~q'}

\label{tab:sos-rules}
\end{table}

The Plotkin style rules for the language $S$ are given in
Table~\ref{tab:sos-rules}; comments on these rules may be found
in~\cite[Sect.~2.1]{He88}. 
\begin{defi}\label{Def:traces} 
  For a sequence of events $\sigma=e_1\cdots e_k$ ($k\geq 0$), and
  states $s,s'$, we write $s \mv{\sigma} s'$ iff there exists a
  sequence of transitions
  \[
   s=s_0 \mv{e_1} s_1 \mv{e_2} \cdots \mv{e_k} s_k=s' \enspace . 
  \]
  If $s \mv{\sigma} s'$ holds for some state $s'$, then $\sigma$ is a {\em
    trace} of $s$. 
  
  The {\sl depth} of a state $s$, written $\depth(s)$, is the length of
  the longest trace it affords.
\end{defi}
In this paper, we shall consider the language $\HCCS$, and more
generally the set of states $S$, modulo split-2 bisimulation
equivalence~\cite{AH93,GV87,GorrieriLaneve1995,He88}. (The weak
variant of this relation is called {\sl t-observational equivalence}
by Hennessy in~\cite{He88}.  Later on, this relation has been called
{\sl timed equivalence} in~\cite{AH93}. Here we adopt the terminology
introduced by van Glabbeek and Vaandrager in~\cite{GV87}.)
\begin{defi}\label{Def:bisimulation}
  {\sl Split-2 bisimulation equivalence}, denoted by $\bistext$, is
  the largest symmetric relation over $S$ such that whenever $s \bis
  t$ and $s \mv{e} s'$, then there is a transition $t \mv{e} t'$ with
  $s' \bis t'$. 

  We shall also sometimes refer to $\bistext$ as {\sl
    split-2 bisimilarity}.   
  If $s \bis t$, then we say that $s$ and $t$ are {\sl split-2
    bisimilar}.
\end{defi}
In what follows, we shall mainly be interested in $\bistext$ as it
applies to the language $\HCCS$. The interested reader is referred
to~\cite[Sect.~2.1]{He88} for examples of (in)equivalent terms with
respect to $\bistext$. Here, we limit ourselves to remarking that
$\bistext$ is a non-interleaving equivalence. For example, the reader
can easily check that the three terms $a~|~b$, $a~ |~b + ab$ and $ab +
ba$ are pairwise inequivalent.

It is well-known that split-2 bisimulation equivalence is indeed an
equivalence relation.  Moreover, two split-2 bisimulation equivalent
states afford the same finite non-empty set of traces, and have
therefore the same depth. 

The following result can be shown following standard lines---see,
e.g.,~\cite{AH93}.
\begin{fact}\label{Fact:congruence}
  Split-2 bisimilarity is a congruence over the language
  $\HCCS$. Moreover, for all states $s,s',t,t'$, if $s \bis s'$ and
  $t\bis t'$, then $s~|~t \bis s'~|~t'$.
\end{fact}
A standard question a process algebraist would ask at this point, and
the one that we shall address in the remainder of this paper, is
whether split-2 bisimulation equivalence affords a finite equational
axiomatization over the language $\HCCS$.  As we showed
in~\cite{AFIL2003}, standard bisimulation equivalence is not finitely
based over the language $\HCCS$. In particular, we argued there 
that no finite collection of equations over $\HCCS$ that is sound with
respect to bisimulation equivalence can prove all of the equations
  \begin{eqnarray}\label{Eqn:family}
  e_n: \quad a\nil \hmerge p_n & \approx & 
  a p_n + \sum_{i=0}^{n} \tau a^i \quad (n\geq 0) \enspace ,
  \end{eqnarray}
  where $a^0$ denotes {\bf 0}, $a^{m+1}$ denotes $a a^m$, and the
  terms $p_n$ are defined thus:
  \begin{eqnarray*}
  p_n & = &  \sum_{i=0}^{n} \bar{a} a^i \quad (n\geq 0) \enspace .
  \end{eqnarray*}
  Note, however, that none of the equations $e_n$ holds with respect
  to $\bistext$. In fact, for each $n\geq  0$, the transition
\[
a p_n + \sum_{i=0}^{n} \tau a^i \mv{S(a)} a_S p_n 
\]
cannot be matched, modulo $\bistext$, by the term $a\nil \hmerge p_n$.
Indeed, the only state reachable from $a\nil \hmerge p_n$ via an
$S(a)$-labelled transition is $a_S \nil ~|~ p_n$. This state is not
split-2 bisimilar to $a_S p_n$ because it can perform the transition
\[
a_S \nil ~|~ p_n \mv{S(\bar{a})} a_S \nil ~|~ \bar{a}_S \nil \enspace , 
\]
whereas the only initial event $a_S p_n$ can embark in is $F(a)$. Thus
the family of equations on which our proof of the main result
from~\cite{AFIL2003} was based is unsound with respect to split-2
bisimilarity. Indeed, as we shall show in what follows, split-2
bisimilarity affords a finite equational axiomatization over the
language $\HCCS$, assuming that the set of actions $\Act$ is
finite. Hence it is possible to finitely axiomatize split-2
bisimilarity over CCS using a single auxiliary binary operation,
viz.~Hennessy's merge.

\section{An Axiomatization of Split-2 Bisimilarity over $\HCCS$}\label{Sect:ax}

Let $\mathcal{E}$ denote the collection of equations 
in Table~\ref{tab:hccs}. In those equations the symbols
$x,y,w,z$ are variables. 
Equation HM6 is
an axiom schema describing one equation per visible action $a$. Note
that $\mathcal{E}$ is finite, if so is $\Act$.

\begin{table}
\centering
\caption{\label{tab:hccs}The Axiom System $\mathcal{E}$ for $\HCCS$ Modulo $\bistext$}
$
\begin{array}{|rrcl|}
\hline
&&&\\
~~~~~ {\rm A}1~~&x+y &\approx& y+x\\
{\rm A}2~~&(x+y)+z &\approx& x+(y+z)~~~~~\\
{\rm A}3~~&x+x &\approx& x\\
{\rm A}4~~&x+{\mathbf 0} &\approx& x\\
{\rm HM}1~~ & (x+y) \hmerge z & \approx & x \hmerge z + y \hmerge z \\

{\rm HM}2~~ & (x \hmerge y) \hmerge z & \approx & x \hmerge (y~|~z) \\
{\rm HM}3~~ & x \hmerge \nil &\approx& x \\
{\rm HM}4~~ & \nil \hmerge x & \approx& \nil \\
{\rm HM}5~~ & (\tau x) \hmerge y & \approx& \tau (x~|~y) \\ 
{\rm HM}6~~ & a x \hmerge ((\bar{a}y \hmerge w)+z) & \approx& a x \hmerge ((\bar{a}y \hmerge w)+z) + \tau (x~|~y ~|~w) \\
{\rm M}~~& x~|~y & \approx & (x\hmerge y) + (y \hmerge x )\\
&&&\\
\hline
\end{array}
$
\end{table}
We write $\mathcal{E} \vdash p \approx q$, where $p,q$ are terms in
the language $\HCCS$ that may possibly contain occurrences of
variables, if the equation $p \approx q$ can be proven from those in
$\mathcal{E}$ using the standard rules of equational logic. For
example, using axioms A1, A2, A4, M, HM1, HM2, HM3 and HM4, it is possible to
derive the equations:
\begin{eqnarray}
\label{Eqn:identity1} x ~|~ \nil & \approx & x \\
\label{Eqn:identity2} \nil~|~ x & \approx & x \\
\label{Eqn:comm} x~|~y & \approx & y~|~ x 
\quad \text{and}\\
\label{Eqn:ass} (x~|~y)~|~z & \approx & x~|~(y~|~z) 
\end{eqnarray}
that state that, modulo $\bistext$, the language $\HCCS$ is a
commutative monoid with respect to parallel composition with $\nil$ as
unit element.
(In light of the provability of
(\ref{Eqn:ass}), we have taken the liberty of omitting parentheses in
the second summand of the term at the right-hand side of equation HM6
in Table~\ref{tab:hccs}.) Moreover, it is easy to see that:
\begin{fact}\label{Fact:nil}
  For each $\HCCS$ term $p$, if $p\bis \nil$, then the equation
  $p\approx \nil$ is provable using A4, HM4 and M.
\end{fact}
All of the equations in the axiom system $\mathcal{E}$ may be found in
the axiomatization of t-observational congruence proposed by Hennessy
in~\cite{He88}. However, the abstraction from $\tau$-labelled
transitions underlying t-observational congruence renders axiom HM2
above unsound modulo that congruence.  (See the discussion
in~\cite[Page~854 and Sect.~3]{Ac94c}.)  Indeed, to the best of our
knowledge, it is yet unknown whether (t-)observational congruence
affords a finite equational axiomatization over CCS, with or without
Hennessy's merge.

Our aim, in the remainder of this note, will be to show that, in the
presence of a finite collection of actions $\Act$, split-2
bisimilarity  {\sl is} finitely
axiomatizable over the language $\HCCS$. This is the import of the
following:
\begin{thm}\label{Thm:main-result}
  For all $\HCCS$ terms $p,q$ not containing occurrences of variables,
  $p \bis q$ if, and only if, $\mathcal{E}\vdash p\approx q$.
\end{thm}
We now proceed to prove the above theorem by establishing separately
that the axiom system $\mathcal{E}$ is sound and complete.
\begin{prop}[Soundness]\label{Prop:soundness}
  For all $\HCCS$ terms $p,q$, if $\mathcal{E}\vdash p\approx q$, then
  $p \bis q$.
\end{prop}
\proof
  Since $\bis$ is a congruence over the language $\HCCS$
  (Fact~\ref{Fact:congruence}), it suffices only to check that each of
  the equations in $\mathcal{E}$ is sound. The verification is
  tedious, but not hard, and we omit the details. 
\qed
\begin{rem}\label{Rem:states}
  For later use in the proof of
  Proposition~\ref{Prop:unique-factorization}, we note that equations
  (\ref{Eqn:identity1})--(\ref{Eqn:ass}) also hold modulo $\bistext$
  when the variables $x,y,z$ are allowed to range over the set of
  states $S$.
\end{rem}
The proof of the completeness of the equations in $\mathcal{E}$ with
respect to $\bis$ follows the general outline of that
of~\cite[Theorem~2.1.2]{He88}. As usual, we rely upon the existence of
normal forms for $\HCCS$ terms. In the remainder of this paper,
process terms are considered modulo associativity and commutativity of
$+$. In other words, we do not distinguish $p+q$ and $q+p$, nor
$(p+q)+r$ and $p+(q+r)$. This is justified because, as previously
observed, split-2 bisimulation equivalence satisfies axioms A1, A2 in
Table~\ref{tab:hccs}.  In what follows, the symbol $=$ will denote
equality modulo axioms A1, A2. We use a {\em summation}
$\sum_{i\in\{1,\ldots,k\}}p_i$ to denote $p_1+\cdots+p_k$, where the
empty sum represents {\bf 0}.
\begin{defi}\label{Def:nf}
  The set $\text{NF}$ of {\sl normal forms} is the least subset of
  $\HCCS$ such that
  \[
  \sum_{i\in I} (a_i p_i \hmerge p_i') + \sum_{j\in J} \tau q_j \in \text{NF} 
  \enspace ,
  \]
  where $I,J$ are finite index sets, if the following conditions hold: 
  \begin{enumerate}
    \item the terms $p_i, p_i'$ ($i\in I$) and $q_j$ ($j\in J$) are
      contained in $\text{NF}$ and
    \item \label{tau-closure} if $a_i p_i \hmerge p_i' \mv{\tau} q$ for
      some $q$, then $q = q_j$ for some $j\in J$.
  \end{enumerate}
\end{defi}
\begin{prop}[Normalization]\label{Prop:nf}
  For each $\HCCS$ term $p$, there is a term $\hat{p}\in\text{NF}$
  such that $\mathcal{E}\vdash p \approx \hat{p}$.  
\end{prop}
\proof
  Define the relation $\sqsubset$ on $\HCCS$ terms thus:
  \begin{quotation}\noindent
    $p \sqsubset q$ if, and only if,
    \begin{itemize}
      \item $\depth(p)<\depth(q)$ or
      \item $\depth(p)=\depth(q)$ and the size of $p$ is smaller than
        that of $q$.
    \end{itemize}
  \end{quotation}
  Note that $\sqsubset$ is a well-founded relation, so we may use
  $\sqsubset$-induction. The remainder of the proof consists of a case
  analysis on the syntactic form of $p$.
  
  We only provide the details for the case $p=q\hmerge r$. (The
  cases $p=\nil$, $p=q+r$ and $p=\mu q$ are trivial---the last owing to
  the fact that $\mu q \approx \mu q\hmerge \nil$ is an instance of
  axiom HM3---, and the case $p=q~|~r$ follows from the case that is
  treated in detail using axiom M.)
  
  Assume therefore that $p=q\hmerge r$. Then $\depth(q)\leq\depth(p)$
  and the size of $q$ is smaller than that of $p$, so $q\sqsubset p$.
  Hence, by the induction hypothesis there exists
  $\hat{q}\in\mathrm{NF}$ such that $\mathcal{E}\vdash
  q\approx\hat{q}$, say
  \begin{equation*}
    \hat{q}=
      \sum_{i\in I}(a_iq_i\hmerge q_i')
        +
      \sum_{j\in J}\tau q_j''
  \enskip.
  \end{equation*}
  By axioms $\mathrm{HM}1$, $\mathrm{HM}2$, $\mathrm{HM}4$ and
  $\mathrm{HM}5$ it follows that
  \begin{equation*}
    p \approx
      \sum_{i\in I}a_iq_i\hmerge(q_i' \mathbin{|} r)
        +
      \sum_{j\in J}\tau(q_j'' \mathbin{|} r)
  \enskip.
  \end{equation*}
  Since $\depth(q_i'),\depth(q_j'') < \depth(\hat{q}) = \depth(q)$ for
  each $i\in I$ and $j\in J$, it follows that
  \[
   \depth(q_i'\mathbin{|}r),\depth(q_j''\mathbin{|}r)
       <\depth(\hat{q}\hmerge r)=\depth(q\hmerge r)=\depth(p) \enspace ,
  \]
  and hence $q_i'\mathbin{|}r\sqsubset p$ and $q_j''\mathbin{|}r\sqsubset p$.
  By the induction hypothesis there are normal forms $\widehat{q_i'\mathbin{|}r}$,
  $\widehat{q_j''\mathbin{|}r}$ such that
    $\mathcal{E}\vdash q_i'\mathbin{|} r\approx \widehat{q_i'\mathbin{|}r},\
                       q_j''\mathbin{|} r\approx \widehat{q_j''\mathbin{|}r}$. 
  So $\mathcal{E}$ proves the equation 
  \begin{equation}\label{Eqn:temp}
    p \approx
      \sum_{i\in I}a_iq_i\hmerge (\widehat{q_i'\mathbin{|}r})
        +
      \sum_{j\in J}\tau (\widehat{q_j''\mathbin{|}r})
  \enskip.
  \end{equation}
  Finally, using equation HM6, it is now a simple matter to add
  summands to the right-hand side of the above equation in order to
  meet requirement~\ref{tau-closure} in Definition~\ref{Def:nf}. In
  fact, let $i\in I$ and 
    \begin{equation*}
    \widehat{q_i'\mathbin{|}r} = \sum_{h\in H}({a_h}r_h\hmerge r_h')
           +
      \sum_{k\in K}\tau r_k''
  \enskip.
  \end{equation*}
  Using A4, we have that 
  \[
  \widehat{q_i'\mathbin{|}r} \approx  \sum_{h\in H,
    a_h=\bar{a_i}}({a_h}r_h\hmerge r_h') + \sum_{h\in H,
    a_h\neq\bar{a_i}}({a_h}r_h\hmerge r_h') + \sum_{k\in K}\tau r_k''
  \]
  is provable from $\mathcal{E}$. 
  Then, using HM6 and the induction hypothesis repeatedly, we can
  prove the equation
  \begin{eqnarray*}
  a_iq_i\hmerge (\widehat{q_i'\mathbin{|}r}) & \approx & 
a_iq_i\hmerge (\widehat{q_i'\mathbin{|}r}) + \sum_{h\in H,
    a_h=\bar{a_i}} \tau \widehat{(q_i \mathbin{|} r_h \mathbin{|} r'_h)} \enspace . 
  \end{eqnarray*}
  Using this equation as a rewrite rule from left to right in
  (\ref{Eqn:temp}) for each $i\in I$ produces a term meeting
  requirement~\ref{tau-closure} in Definition~\ref{Def:nf} that is the
  desired normal form for $p=q \hmerge r$.  
\qed
The key to the proof of the promised completeness theorem is an
important cancellation result that has its roots in one proven by
Hennessy for his t-observational equivalence in~\cite{He88}. 
\begin{thm}\label{Thm:decomposition}
  Let $p,p',q,q'$ be $\HCCS$ terms, and let $a$ be a visible action.
  Assume that 
  \[
  a_S p ~|~p' \bis a_S q ~|~q' \enspace .
  \] 
  Then $p\bis q$ and $p'\bis q'$.
\end{thm}
For the moment, we postpone the proof of this result, and use it to
establish the following statement, to the effect that the axiom system
$\mathcal{E}$ is complete with respect to $\bis$ over $\HCCS$.
\begin{thm}[Completeness]\label{Thm:completeness}
  Let $p,q$ be $\HCCS$ terms such that $p \bis q$. Then
  $\mathcal{E}\vdash p \approx q$.
\end{thm}
\proof
  By induction on the depth of $p$ and $q$. (Recall that, since $p
  \bis q$, the terms $p$ and $q$ have the same depth.) In light of
  Proposition~\ref{Prop:nf}, we
  may assume without loss of generality that $p$ and $q$ are contained
  in $\text{NF}$. Let 
  \begin{eqnarray*}
  p & = & \sum_{i\in I} (a_i p_i \hmerge p_i') + \sum_{j\in J} \tau p_j'' \quad \text{and}\\
  q & = & \sum_{h\in H} (b_h q_h \hmerge q_h') + \sum_{k\in K} \tau q_k''\enspace .
  \end{eqnarray*}
  We prove that $\mathcal{E} \vdash p \approx p+q$, from which the
  statement of the theorem follows by symmetry and transitivity. To
  this end, we argue that each summand of $q$ can be absorbed into $p$
  using the equations in $\mathcal{E}$, i.e., that
  \begin{enumerate}
    \item \label{tauabs} $\mathcal{E} \vdash p \approx p + \tau q_k''$ for each $k\in K$, and 
    \item \label{actabs} $\mathcal{E} \vdash p \approx p + (b_h q_h \hmerge q_h')$ for each $h\in H$. 
  \end{enumerate}
  We prove these two statements in turn. 
  \begin{itemize}
\item {\sc Proof of Statement~\ref{tauabs}}. Let $k\in K$. Then
    $q\mv{\tau} q_k''$. Since $p\bis q$, there is a term $r$ such that
    $p \mv{\tau} r$ and $r \bis q_k''$. Since $p\in \text{NF}$,
    condition~\ref{tau-closure} in Definition~\ref{Def:nf} yields that
    $r=p''_j$ for some $j\in J$. The induction hypothesis together
    with closure with respect to $\tau$-prefixing now yields that
    \[
    \mathcal{E} \vdash \tau p_j'' \approx \tau q_k'' \enspace .
    \]
    Therefore, using A1--A3, we have that 
    \[
    \mathcal{E} \vdash p \approx p + \tau p_j'' \approx p + \tau q_k'' \enspace ,
    \]
    which was to be shown. 
  \item {\sc Proof of Statement~\ref{actabs}}. Let $h\in H$. Then
    $q\mv{S(b_h)} {b_h}_S q_h ~|~ q_h'$. Since $p\bis q$, there is a
    state $s$ such that $p \mv{S(b_h)} s$ and $s \bis {b_h}_S q_h ~|~
    q_h'$. Because of the form of $p$, it follows that $s = {a_i}_S
    p_i ~|~ p_i'$ for some $i\in I$ such that $a_i=b_h$. By
    Theorem~\ref{Thm:decomposition}, we have that 
    \[
    p_i \bis q_h \text{ and } p_i' \bis q_h' \enspace .
    \]
    Since the depth of all of these terms is smaller than that of $p$,
    we may apply the induction hypothesis twice to obtain that 
    \[
    \mathcal{E} \vdash p_i \approx q_h \text{ and } \mathcal{E} \vdash
    p_i' \approx q_h' \enspace .
    \]
    Therefore, using A1--A3 and $a_i = b_h$, we have that 
    \[
    \mathcal{E} \vdash p \approx p + (a_i p_i \hmerge p_i') \approx p + (b_h q_h \hmerge q_h') \enspace ,
    \]
    which was to be shown.     
  \end{itemize}
  The proof of the theorem is now complete. 
\qed
To finish the proof of the completeness theorem, and therefore of
Theorem~\ref{Thm:main-result}, we are left to show
Theorem~\ref{Thm:decomposition}. Our proof of that result relies on a
unique decomposition property with respect to parallel composition for
states modulo $\bistext$. In order to formulate this decomposition
property, we shall make use of some notions from~\cite{MM93,Mo89}.
These we now proceed to introduce for the sake of completeness and
readability.
\begin{defi}\label{Def:prime}
  A state $s$ is {\sl irreducible} if $s \bis s_1~|~s_2$ implies
  $s_1\bis \nil$ or $s_2\bis \nil$, for all states $s_1,s_2$.
  
  We say that $s$ is {\sl prime} if it is irreducible and is not
  split-2 bisimilar to $\nil$.
\end{defi}
For example, each state $s$ of depth $1$ is prime because every state
of the form $s_1~|~s_2$, where $s_1$ and $s_2$ are not split-2
bisimilar to $\nil$, has depth at least $2$, and thus cannot be split-2
bisimilar to $s$. 
\begin{fact}\label{Fact:aSp-prime}
  The state $a_S p$ is prime, for each $\HCCS$ term $p$ and action $a$.
\end{fact}
\proof
  Since $a_S p$ is not split-2 bisimilar to $\nil$, it suffices only
  to show that it is irreducible. To this end, assume, towards a
  contradiction, that $a_S p \bis s_1~|~s_2$ for some states $s_1,
  s_2$ that are not split-2 bisimilar to $\nil$. Then, since $a_S p
  \bis s_1~|~s_2$, we have that $s_1 \mv{F(a)}s_1'$ and $s_2
  \mv{F(a)}s_2'$, for some $s_1',s_2'$.  But then it follows that
    \[
     s_1~|~s_2 \mv{F(a)}s_1'~|~s_2 \mv{F(a)}s_1'~|~s_2'\enspace ,
    \]
    whereas the term $a_S p$ cannot perform two subsequent
    $F(a)$-transitions.  
    We may therefore conclude that such states $s_1$ and $s_2$ cannot
    exist, and hence that the term $a_S p$ is irreducible, which was
    to be shown.
\qed
The following result is the counterpart for the language $\HCCS$ of
the unique decomposition theorems presented for various languages
in, e.g.,~\cite{AH93,Luttik2003,MM93,Mo89}.

\begin{prop}\label{Prop:unique-factorization}
  Each state is split-2 bisimilar to a parallel composition of primes,
  uniquely determined up to split-2 bisimilarity and the order of the
  primes. (We adopt the convention that $\nil$ denotes the empty
  parallel composition.)
\end{prop}
\proof
  \newcommand{\eqclass}[1]{\ensuremath{[#1]}}
  \newcommand{\merge}{\ensuremath{\mathbin{|}}}
  \newcommand{\ord}{\ensuremath{\mathrel{\preccurlyeq}}}
  \newcommand{\sord}{\ensuremath{\mathrel{\preccurly}}}
  \newcommand{\bisimilar}[1][]{%
    \setbox0=\hbox{\kern-.1ex{$\leftrightarrow$}\kern-.1ex}
    \setbox1=\vbox{\hbox{\raise .1ex \box0}\hrule}%
    \ensuremath{\mathrel{\hbox{\kern.1ex\box1\kern.1ex}_{#1}}}
  }
 
\newcommand{\splbisimilar}{\ensuremath{\mathrel{\bisimilar_{\text{{2S}}}}}}
  
  We shall obtain this result as a consequence of a general unique
  decomposition result, obtained by the fourth author in~\cite{Luttik2003}.

  Let $\eqclass{S}$ denote the set of states modulo split-2
  bisimilarity, and, for a state $s\in S$, denote by $\eqclass{s}$ the
  equivalence class in $\eqclass{S}$ that contains $s$.
  By Fact~\ref{Fact:congruence} we can define on $\eqclass{S}$ a binary
  operation $\merge$ by
  \begin{equation*}
    \eqclass{s}\merge\eqclass{t}
      = \eqclass{s\merge t}
  \enskip.
  \end{equation*}
  By Remark~\ref{Rem:states}, 
  the set $\eqclass{S}$ with the binary operation $\merge$ and the
  distinguished element $\eqclass{\nil}$ is a commutative monoid.

  Next, we define on $\eqclass{S}$ a partial order $\ord$ by
  \begin{equation*}
    \eqclass{s'}\ord\eqclass{s}\
  \text{iff}\
           \text{there exist $s''\in S$ and $\sigma\in E^*$
      such that $s\goto{\sigma}s''\bis s'$}.
  \end{equation*}
  Note that $\ord$ is indeed a partial order (to establish
  antisymmetry use that transitions decrease depth,  and that split-2
  bisimilar states have the same depth).

  For each state $s$, there are a sequence of events $\sigma$ and a state $s'$ such that 
  \[
  s \goto{\sigma} s' \bis \nil \enspace .
  \]
  So $\eqclass{\nil}$ is the least element of $\eqclass{S}$ with
  respect to $\ord$.  Furthermore, if $\eqclass{s'}\ord\eqclass{s}$,
  then $s \goto{\sigma} s''\bis s'$, for some $\sigma\in E^*$ and
  state $s''$.  So, using the SOS rules for $S$ and
  Fact~\ref{Fact:congruence}, it follows that 
  \[
  s\merge t\goto{\sigma}
  s''\merge t\bis s'\merge t \enspace , 
  \]
  and hence
  \[
   \eqclass{s'}\merge\eqclass{t}
       =\eqclass{s'\merge t}
       \ord\eqclass{s\merge t}
       =\eqclass{s}\merge\eqclass{t} \enspace .
  \]
  Thereby, we have now established that $\eqclass{S}$ with $\merge$,
  $\eqclass{\nil}$ and $\ord$ is a positively ordered commutative
  monoid in the sense of~\cite{Luttik2003}.
  
  From the SOS rules for $S$ it easily follows that this positively
  ordered commutative monoid is \emph{precompositional}
  (see~\cite{Luttik2003}), i.e., that
  \begin{equation*}
    \text{if $\eqclass{s}\ord\eqclass{s_1}\merge\eqclass{s_2}$,
          then there are
            $\eqclass{s_1'}\ord\eqclass{s_1}$,
            $\eqclass{s_2'}\ord\eqclass{s_2}$
          s.t.\
            $\eqclass{s}=\eqclass{s_1'}\merge\eqclass{s_2'}$}.         
  \end{equation*}
  Consider the mapping $|\_|:\eqclass{S}\rightarrow\mathbf{N}$ into
  the positively ordered monoid of natural numbers with addition, $0$
  and the standard less-than-or-equal relation, defined by
  \begin{equation*}
    \eqclass{s}\mapsto\depth(s)
  \enskip.
  \end{equation*}
  It is straightforward to verify that $|\_|$ is a \emph{stratification}
  (see~\cite{Luttik2003}), i.e., that 
  \begin{enumerate}
  \renewcommand{\theenumi}{\roman{enumi}}
  \renewcommand{\labelenumi}{(\theenumi)}
    \item
      $|\eqclass{s}\merge\eqclass{t}|=|\eqclass{s}|+|\eqclass{t}|$;
  and
    \item
      if $\eqclass{s}\prec\eqclass{t}$, then
$|\eqclass{s}|<|\eqclass{t}|$.
  \end{enumerate}
  We conclude that $\eqclass{S}$ with $\merge$, $\eqclass{\nil}$ and
  $\ord$ is a stratified and precompositional positively ordered
  commutative monoid, and hence, by Theorem~13 in \cite{Luttik2003}, it has
  unique decomposition. This completes the proof of the proposition.
\qed
Using the above unique decomposition result, we are now in a position
to complete the proof of Theorem~\ref{Thm:decomposition}. 
\begin{proofofthm}{Thm:decomposition}
Assume that $a_S p ~|~ p' \bis a_S q ~|~q'$. Using
Proposition~\ref{Prop:unique-factorization}, we have that $p'$ and
$q'$ can be expressed uniquely as parallel compositions of primes. Say
that 
\begin{eqnarray*}
p' & \bis & p_1 ~|~ p_2 ~|~ \cdots ~|~ p_m \quad \text{and}\\
q' & \bis & q_1 ~|~ q_2 ~|~ \cdots ~|~ q_n 
\end{eqnarray*}
for some $m,n\geq 0$ and primes $p_i$ ($1\leq i \leq m$) and $q_j$
($1\leq j \leq n$) in the language $\HCCS$. Since $a_S p$ and $a_S q$
are prime (Fact~\ref{Fact:aSp-prime}) and $\bistext$ is a congruence
(Fact~\ref{Fact:congruence}), the unique prime decompositions of $a_S
p~|~p'$ and $a_S q~|~q'$ given by
Proposition~\ref{Prop:unique-factorization} are
\begin{eqnarray*}
a_S p~|~p & \bis & a_S p ~|~ p_1 ~|~ p_2 ~|~ \cdots ~|~ p_m \quad \text{and}\\
a_S q~|~ q' & \bis & a_S q~|~ q_1 ~|~ q_2 ~|~ \cdots ~|~ q_n \enspace ,
\end{eqnarray*}
respectively. In light of our assumption that $a_S p ~|~ p' \bis a_S q
~|~q'$, these two prime decompositions coincide by
Proposition~\ref{Prop:unique-factorization}. Hence, as for each $1\leq j \leq n$\begin{eqnarray*}
a_S p & \nbis & \quad q_j \enspace , 
\end{eqnarray*}
we have that
\begin{enumerate}
\item $a_S p\bis a_S q$, 
\item $m=n$ and, without loss of generality, 
\item $p_i \bis q_i$ for
  each $1\leq i \leq m$.
\end{enumerate}
It is now immediate to see that $p \bis q$ and $p' \bis q'$, which was
to be shown.
\end{proofofthm}

\section*{Acknowledgements}
We thank the referees for suggestions that led to improvements in the
presentation of the paper. The work reported in this paper was carried
out while Luca Aceto was on leave at Reykjav\'{\i}k University, Wan
Fokkink was at CWI, and Anna Ing\'olfsd\'ottir was at deCODE
Genetics. They thank these institutions for their hospitality and
excellent working conditions.  Luca Aceto's work was partially
supported by the Statens Naturvidenskabelige Forskningsr{\aa}d (Danish
Natural Science Research Council), project ``The Equational Logic of
Parallel Processes'', nr.~21-03-0342.

%


\end{document}